\documentclass[conference]{IEEEtran}
\IEEEoverridecommandlockouts

\usepackage{cite}
\usepackage{amsmath,amssymb,amsfonts}
\usepackage{algorithmic}
\usepackage{graphicx}
\usepackage{textcomp}
\usepackage{xcolor}
\usepackage[caption=false,font=footnotesize]{subfig}

\usepackage{booktabs}
\usepackage{enumitem}
\usepackage[pagebackref,breaklinks,colorlinks]{hyperref}

\usepackage[capitalize]{cleveref}
\crefname{section}{Sec.}{Secs.}
\Crefname{section}{Section}{Sections}
\Crefname{table}{Table}{Tables}
\crefname{table}{Tab.}{Tabs.}

\def\BibTeX{{\rm B\kern-.05em{\sc i\kern-.025em b}\kern-.08em
    T\kern-.1667em\lower.7ex\hbox{E}\kern-.125emX}}
\begin{document}

\title{MEGANet-W: A Wavelet-Driven Edge-Guided Attention Framework for Weak Boundary Polyp Detection}

\author{%
  \IEEEauthorblockN{Tan Zhe Yee}
  \IEEEauthorblockA{%
    School of Artificial Intelligence and Robotics (SAIR)\\
    Xiamen University Malaysia, Sepang, Malaysia\\
    ait2209953@xmu.edu.my}%
    \and
    \IEEEauthorblockN{Ashwaq Qasem\textsuperscript{*}}
  \IEEEauthorblockA{%
    School of Artificial Intelligence and Robotics (SAIR)\\
    Xiamen University Malaysia, Sepang, Malaysia\\
    ashwaq.qasem@xmu.edu.my}%
}

\maketitle

\begin{abstract}
Colorectal polyp segmentation is critical for early detection of colorectal cancer, yet weak and low contrast boundaries significantly limit automated accuracy. Existing deep models either blur fine edge details or rely on handcrafted filters that perform poorly under variable imaging conditions. We propose MEGANet-W, a Wavelet Driven Edge Guided Attention Network that injects directional, parameter free Haar wavelet edge maps into each decoder stage to recalibrate semantic features. The key novelties of MEGANet-W include a two-level Haar wavelet head for multi orientation edge extraction; and Wavelet Edge Guided Attention (W-EGA) modules that fuse wavelet cues with boundary and input branches. On five public polyp datasets, MEGANet-W consistently outperforms existing methods, improving mIoU by up to 2.3\% and mDice by 1.2\%, while introducing no additional learnable parameters. This approach improves reliability in difficult cases and offers a robust solution for medical image segmentation tasks requiring precise boundary detection.

\end{abstract}


\section{Introduction}
\begin{figure}[t]
  \centering
  \subfloat[]{\includegraphics[width=0.30\columnwidth]{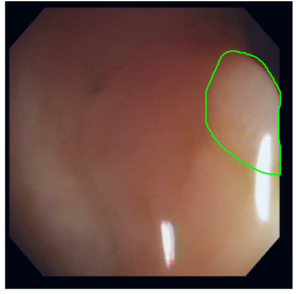}}\hfill
  \subfloat[]{\includegraphics[width=0.30\columnwidth]{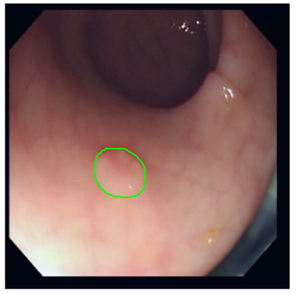}}\hfill
  \subfloat[]{\includegraphics[width=0.30\columnwidth]{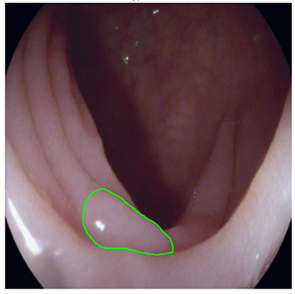}}\\[1ex]

    \makebox[\columnwidth][c]{%
      \subfloat[]{\includegraphics[width=0.30\columnwidth]{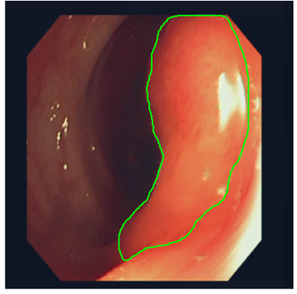}}%
      \quad
      \subfloat[]{\includegraphics[width=0.30\columnwidth]{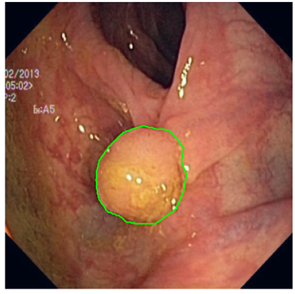}}%
    }

  \caption{Representative colorectal-polyp examples from five public datasets, with the polyp region highlighted by a green outline: (a) CVC-ColonDB; (b) CVC-300; (c) ETIS-LaribPolypDB; (d) CVC-ClinicDB; (e) Kvasir.}
  \label{fig:polyp-examples}
\end{figure}

\noindent The colorectal cancer (CRC) is one of the leading common cancers all over the globe and has a cancer mortality ranking of third highest \cite{Wang2025}. Early prevention is essential for saving public health, and colonoscopy is widely regarded as the "gold standard" for detecting polyps, which play an important role in the prevention of progression into CRC \cite{Li2025}. However, the effectiveness of colonoscopy highly depends on clinician’s skills, with a 26\% missed detection when put through a manual screening procedure \cite{Quan2022, Wang2025}. As illustrated in Fig.~\ref{fig:polyp-examples}, accurate segmentation of polyps remains a challenging task due to the high visual similarity between polyps and surrounding normal mucosal tissues in terms of color and texture \cite{Tong2025}. Such inherent ambiguity complicates semantic representation and makes precise localization of polyps even difficult \cite{Wang2025}. Hence, improving the polyps detection accuracy is importance in reducing risk of developing colorectal cancer \cite{Bui2023}.\\

\noindent Recent advances in deep learning (DL) has shown big potential in helping doctors to more accurately localize the polyp \cite{Li2025}. In contrast to DL methods, traditional approaches rely on handcrafted features, like edge detection or threshold-based segmentation \cite{Wu2024}. However, these techniques are very sensitive to different in lighting and show limited feature representation capabilities, which restrict their applicability in clinical settings \cite{Mei2025, Quan2022}. DL-based methods on the other hand, are able to learn rich and hierarchical feature representations directly from the image data, increase the applicability to diverse imaging conditions and effectively overcoming the limitations of handcrafted techniques \cite{Mei2025, Wu2024}.\\

\noindent U-Net along with variants, which are based on convolutional neural networks, have facilitated the rapid development of medical image segmentation tasks \cite{Wang2025}. Subsequent studies use Transformer as encoder to strengthen the network’s global feature representation \cite{Xu2023, Zhang2021}. Attention mechanism like SE-Net \cite{Hu2018} and CBAM \cite{Woo2018} improve feature selectivity but offen blur boundaries due to spatial down-sampling, especially in low-contrast regions. \\

\noindent To enhance edge preservation, recent approaches leverage frequency-domain processing via discrete wavelet transform (DWT). Methods such as FAENet \cite{Tang2025}, WRANet \cite{Zhao2023}, and WBANet \cite{Wang2025} integrate wavelet-based decomposition and attention mechanisms to improve edge sharpness and texture consistency, particularly in polyp segmentation. Beyond wavelet-based models, SFA \cite{FangYuqiandChen2019} uses boundary-aware supervision for edge refinement, while PraNet \cite{Fan2020} applies reverse attention for ambiguous regions. MEGANet \cite{Bui2023} introduces Laplacian-based edge cues but relies on classical edge detection and lacks deep frequency-domain integration. \\

\noindent In this work, we propose \textbf{MEGANet-W}, a Wavelet-Driven Edge-Guided Attention Network that integrates frequency-domain processing into every decoder stage. Our contributions are as follows:

\begin{itemize}
  \item \textbf{Two‐level Haar wavelet edge head:}
    A parameter‐free, directional high‐frequency extractor that
    enhances boundary representation without introducing additional
    trainable weights.

  \item \textbf{Wavelet Edge‐Guided Attention modules:}
    A wavelet‐based extension uses edge cues derived from wavelet transforms at each decoder stage to recalibrate semantic feature maps confidence.
\end{itemize}

\section{Related Work}
\noindent We categorize relevant work into three domains: edge-aware segmentation, wavelet-based feature extraction, and attention mechanisms. Our method builds upon these by integrating wavelet-derived edge features within a decoder attention framework for improved polyp segmentation.\\

\noindent \textbf{Edge-Aware Polyp Segmentation. }Accurate edge preservation is crucial in polyp segmentation due to the subtle yet critical boundaries between polyps and surrounding mucosa \cite{Chen2023}. Early approaches utilized hand-crafted edge detectors such as Canny, Sobel or Laplacian filters, but these suffered from sensitivity to noise and lacked semantic understanding \cite{Isar2021}. Recent deep learning-based approaches integrating boundary awareness directly into segmentation networks \cite{Yue2024}. SFA \cite{FangYuqiandChen2019} enhances boundary modelling through an auxiliary decoder dedicated to edge prediction and a boundary-sensitive loss that enforce area-boundary consistency. PraNet \cite{Fan2020}, introduces reverse attention to enhance segmentation in ambiguous boundary regions. Following this, MEGANet \cite{Bui2023} combines encoder features, decoder predictions, and Laplacian-derived high-frequency maps via a Multi-Scale Edge-Guided Attention (EGA). \cite{Lee2023} further refined boundaries by combining shallow features with reverse attention, while \cite{Lei2024} and \cite{Zhang2024} incorporated dual attention mechanisms tailored for edge modeling. However, these approaches either rely on learned edge labels or hand-crafted filters, and do not leverage the frequency domain explicitly for boundary enhancement.\\

\noindent \textbf{Wavelet-Based Feature Extraction. }Wavelets are effective in extracting localized frequency components and have been used for denoising, texture characterization, and multi-resolution analysis \cite{Vyas2016}. Their incorporation into convolution neural networks has enhanced feature representation, especially in medical imaging tasks \cite{Fujieda2018,Kang2017}. \cite{Li2020} integrated DWT into U-Net like architecture to improve segmentation robustness. \cite{Zhao2023} replaces pooling layers with DWT to better separate noise-prone high-frequency and stable low-frequency components. More recently, \cite{Zeng2024} introduced wavelet pooling in a transformer-based framework to preserve spatial detail, and \cite{El-Khamy2025} proposed matched wavelet pooling (MWP) to further improve semantic segmentation. Despite these advances, most wavelet-based models limit wavelet integration to encoder or pooling stages, without leveraging frequency-aware edge maps in the decoder. Our work explicitly injects wavelet-derived edges into the decoder attention mechanism to better localize polyp boundaries.\\

\noindent \textbf{Attention Mechanisms in Polyp Segmentation. }Attention mechanisms are foundational in modern segmentation architectures, enabling adaptive focus on informative spatial and channel-wise features. Early approaches such as SE-Net \cite{Hu2018} and CBAM \cite{Woo2018} enhanced global feature selectivity. More recent advances have targeted boundary precision in medical image segmentation. MSNet \cite{Zhao2021} introduces a multi-scale subtraction module to suppress redundant information and extract complementary features, improving attention-based segmentation across scales. Polyper \cite{Shao2023} introduces boundary-sensitive attention to distinguish polyp interiors from edges, while LDNet \cite{Zhang2023} employs lesion-aware cross-attention to refine ambiguous contours. PEFNet \cite{Nguyen-Mau2023} integrates positional encoding of polyp regions into skip connections, guiding attention toward structurally salient regions. MISNet \cite{Kang2024} adopts a parallel attention structure with a balancing module to enhance boundary localization across scales.\\

\noindent Meanwhile, frequency-aware attention has gained traction for its ability to separate edge and context features. FAENet \cite{Tang2025} apply DWT to decompose features into high- and low-frequency bands, and WLAM \cite{Feng2025} combines wavelet-derived low- and high- frequency features within a linear attention framework. Although promising, these designs are often limited to encoder-side processing or classification tasks, and few apply frequency-domain signals directly within decoder attention. Most closely related, WBANet \cite{Wang2025} utilizes wavelet-based attention into a boundary-aware segmentation framework, yet lacks explicit edge-map injection into attention queries. Our work bridges this gap by feeding wavelet-derived edge features into decoder attention, enabling the model to adaptively focus on high-frequency boundary information during reconstruction, combining the strengths of spatial attention with the domain-adaptivity of wavelets. 

\section{Proposed Methodology}
\begin{figure*}[!t]
  \centering
  \includegraphics[width=0.75\textwidth,keepaspectratio]{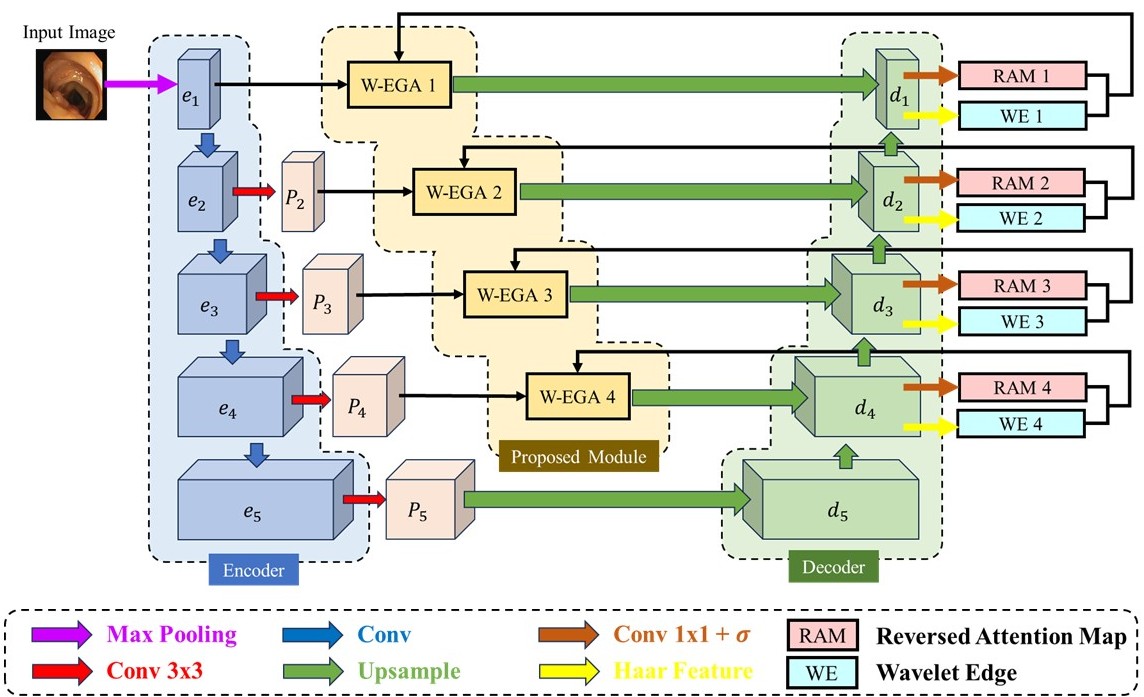}
  \caption{The overall architecture of MEGANet-W includes three modules: the encoder, the decoder (a U-Net to extract visual representations), and the proposed Wavelet Edge-Guided Attention (W-EGA) module.}
  \label{fig:architecture}
\end{figure*}

\noindent This section describes the upgraded MEGANet-W, as depicted in Fig.\ref{fig:architecture}. Compared to the original MEGANet-R34 baseline \cite{Bui2023}, the proposed model: (i) replaces the Laplacian-based edge extractor with a two-level Haar wavelet head, and (ii) integrates the resulting multi-orientation edge features into the decoder via a Wavelet Edge-Guided Attention (W-EGA) block. 

\subsection{Network Architecture}
\label{sec:architecture}

\noindent Our proposed network retains the encoder–decoder architecture of the \cite{Bui2023} while integrating edge information at each decoding stage. Given an RGB image $\mathbf{I}\in\mathbb{R}^{3\times H\times W}$, a ResNet‐34 backbone extracts five spatial‐scale feature maps $\{\mathbf{e}_1,\mathbf{e}_2,\ldots,\mathbf{e}_5\}$ at strides $\{2,4,8,16,32\}$, respectively. Starting from $\mathbf{e}_5$, a symmetric decoder with five upsampling blocks restores the input resolution. At each decoding stage $i\in\{4,3,2,1\}$, except the coarsest, we insert W-EGA module (see Fig.~\ref{fig:architecture}), replacing the Laplacian‐based EGA from \cite{Bui2023}. Each decoder stage $\mathbf{d}_i$ is projected to a side output $\mathbf{P}_i$ via a standard ConvBlock (see Sec.~\ref{sec:w-ega}), followed by a $1\times1$ convolution. The side logits are bilinearly upsampled to the input resolution and supervised jointly.

\subsection{Wavelet Enhanced Edge Head}

\noindent The original MEGANet \cite{Bui2023} detects boundaries using a $5\times5$ Gaussian–Laplacian pyramid, which is isotropic and sensitive to noise. In contrast, we adopt a directional multilevel‐level 2D DWT using orthogonal Haar bases. This enable both fine and coarser boundary cues are being captured without adding learnable parameters. We cascade two fixed $2\times2$ Haar DWTs and fuse their high-frequency sub-bands. Let $\mathbf{X}\in\mathbb{R}^{C\times H\times W}$ denote a decoder feature map at spatial resolution $H\times W$ with $C$ channels.

\noindent \textbf{Level-1 decomposition} applies a standard 2D Haar DWT:
\begin{equation}
\{A^{(1)},\,D^{(1)}_{LH},\,D^{(1)}_{HL},\,D^{(1)}_{HH}\}
= \operatorname{DWT}_{\mathrm{Haar}}(\mathbf{X}),
\label{eq:level1_decomp}
\end{equation}
where 
\begin{equation}
\begin{aligned}
A^{(1)} &\in \mathbb{R}^{\,C \times \tfrac{H}{2} \times \tfrac{W}{2}}, \\[4pt]
D^{(1)}_{XY} &\in \mathbb{R}^{\,C \times \tfrac{H}{2} \times \tfrac{W}{2}}
\quad\text{for }XY\in\{LH,HL,HH\}\,.
\end{aligned}
\label{eq:subband_dims}
\end{equation}
are the low-frequency approximation and the three high-frequency detail sub-bands, respectively.

\noindent \textbf{Level-2 decomposition} is recursively applied to the approximation:
\begin{equation}
\{A^{(2)},\,D^{(2)}_{LH},\,D^{(2)}_{HL},\,D^{(2)}_{HH}\}
= \operatorname{DWT}_{\mathrm{Haar}}\bigl(A^{(1)}\bigr),
\label{eq:level2_decomp}
\end{equation}
yielding a coarser set of detail sub-bands and approximation 

\begin{equation}
A^{(2)} \in \mathbb{R}^{C \times \tfrac{H}{4} \times \tfrac{W}{4}}.
\label{eq:level2_dims}
\end{equation}


\noindent Implementation‐wise, these coefficients are obtained via depth‐wise $2\times2$ convolutions with fixed Haar high‐pass filters:

\begin{equation}
\resizebox{\columnwidth}{!}{%
  $\begin{aligned}
    k_{LH} &= \tfrac12\begin{bmatrix}+1&+1\\-1&-1\end{bmatrix},\quad
    k_{HL} &= \tfrac12\begin{bmatrix}+1&-1\\+1&-1\end{bmatrix},\quad
    k_{HH} &= \tfrac12\begin{bmatrix}+1&-1\\-1&+1\end{bmatrix}.
  \end{aligned}$
}
\end{equation}

\noindent At each level $\ell\in\{1,2\}$, we concatenate the three detail subbands to form
\begin{equation}
W^{(\ell)}
= \bigl[D_{LH}^{(\ell)},\;D_{HL}^{(\ell)},\;D_{HH}^{(\ell)}\bigr]
\;\in\;
\mathbb{R}^{3C \times \frac{H}{2^\ell} \times \frac{W}{2^\ell}}.
\end{equation}
Each tensor $W^{(\ell)}$ is then bilinearly upsampled to the original resolution:
\begin{equation}
\widetilde{W}^{(\ell)}
= \mathrm{Upsample}\bigl(W^{(\ell)},\,\text{scale}=2^\ell\bigr)
\;\in\;
\mathbb{R}^{3C \times H \times W}.
\end{equation}
The final multi-scale edge representation is obtained by summation:
\begin{equation}
\widetilde{W}_{\text{sum}} = \widetilde{W}^{(1)} + \widetilde{W}^{(2)}
\in \mathbb{R}^{3C\times H\times W}.
\end{equation}
This tensor is fed into the subsequent W-EGA modules. The orthonormality of the Haar kernels ensures energy preservation while extracting directional edge bands. This wavelet head operates end-to-end using three fixed 2 × 2 filters, introduces no learnable parameters, and embeds high-frequency priors typically absent from conventional CNN features. The resulting multi-scale edge representation is richer and effectively encodes structural boundaries across multiple resolutions.

\subsection{Wavelet Edge-Guided Attention}
\label{sec:w-ega}
\noindent We redesign the original EGA module \cite{Bui2023} to accommodate our tri‐channel wavelet cues, yielding W‐EGA. At decoder depth $i\le4$, the W‐EGA gate fuses three information sources, all spatially aligned to $H_i\times W_i$:

\paragraph{Reverse‐Attention Branch.}
The sigmoid of the incoming side‐prediction $P_{i+1}$ is inverted to highlight background pixels:
\begin{equation}
\begin{aligned}
R_i &= 1 - \sigma\bigl(P_{i+1}\bigr)
       \;\in\;\mathbb{R}^{B\times1\times H_i\times W_i},\\
F_{\mathrm{bg}} &= F_i \odot R_i.
\end{aligned}
\end{equation}
where $F_i\in\mathbb{R}^{B\times c_i\times H_i\times W_i}$ is the incoming decoder feature and $\sigma(\cdot)$ denotes the sigmoid function. The mask $R_i$ suppresses already‐confident foreground predictions, mirroring the reverse‐attention idea in MEGANet.

\paragraph{Boundary‐Attention Branch.}
Wavelet detail coefficients of the same prediction provide boundary evidence. We collapse wavelet edges of the current probability map across sub‐bands to form a soft contour mask:
\begin{equation}
\begin{aligned}
B_i &= \sum_{c} E_{c}\bigl(\sigma(P_{i+1})\bigr)
       \;\in\;\mathbb{R}^{B\times1\times H_i\times W_i},\\
F_{\mathrm{edge}} &= F_i \odot B_i.
\end{aligned}
\end{equation}

\noindent where $E_c(\cdot)$ extracts the $c$-th wavelet detail channel. The mask $B_i$ highlights boundary regions for enhanced edge guidance.

\begin{figure}[!t]
  \centering
  \subfloat[Raw RGB input \(F_i\)]{%
    \includegraphics[height=1.5in]{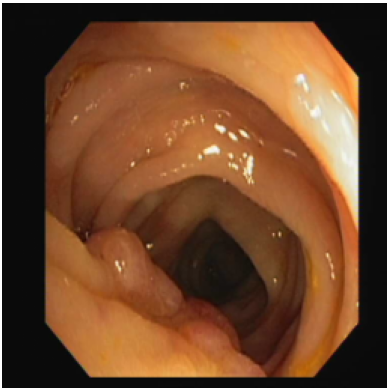}%
    \label{fig:inp-raw}%
  }\hfill
  \subfloat[Edge‐guided input \(F_{\mathrm{inp}}=F_i\odot H_i\)]{%
    \includegraphics[height=1.5in]{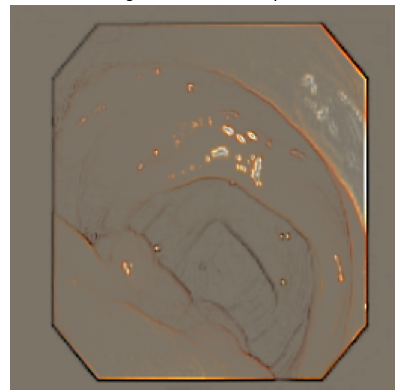}%
    \label{fig:inp-edge}%
  }\\[1ex]
  \subfloat[Detail coeffs (LH, HL, HH) used to build \(H_i\)]{%
    \includegraphics[width=\columnwidth,keepaspectratio]{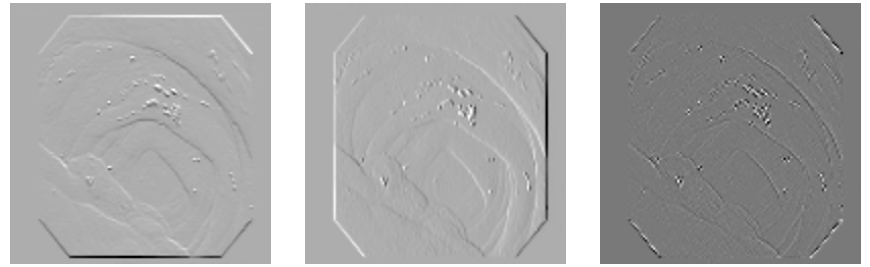}%
    \label{fig:inp-coef}%
  }
  \caption{Visualization of the Input‐Edge Branch.  
    (a) Original colonoscopy frame,  
    (b) channel‐wise wavelet edge map overlaid on RGB,  
    (c) the three high-frequency orientation maps whose absolute sum yields \(H_i\).}
  \label{fig:preproc}
\end{figure}

\paragraph{Input‐Edge Branch.}
Wavelet edges of the grayscale input supply high‐frequency priors, see Fig.\ref{fig:preproc}:
\begin{equation}
\begin{aligned}
H_i &= \sum_{c} E_{c}\bigl(\bar{F}_i\bigr),\\
F_{\mathrm{inp}} &= F_i \odot H_i,
\end{aligned}
\quad
\bar{F}_i = \tfrac{1}{C_i}\sum_{k=1}^{C_i}F_i^{(k)},
\end{equation}
where $\bar{F}_i$ is the channel‐wise mean of $F_i$. This branch injects texture cues independent of the current logits.\\

\noindent The three feature tensors share spatial size $H_i\times W_i$ and channel count $C_i$. They are concatenated and compressed via
\begin{equation}
\begin{aligned}
Z_i &= \phi\bigl([\;F_{\mathrm{bg}};F_{\mathrm{edge}};F_{\mathrm{inp}}\;]\bigr),\\
    &= \mathrm{ReLU}\bigl(\mathrm{BN}(\mathrm{Conv}_{3\times3}(\cdot))\bigr).
\end{aligned}
\end{equation}
where $Z_i\in\mathbb{R}^{B\times C_i\times H_i\times W_i}$. So that $\phi:\,3C_i\!\to\!C_i$. A spatial‐attention squeeze follows:
\begin{equation}
\begin{aligned}
\alpha_i &= \sigma\bigl(\psi(Z_i)\bigr),\\
Z'_i &= Z_i \odot \alpha_i,
\end{aligned}
\end{equation}
where $\psi:\mathbb{R}^{C_i}\to\mathbb{R}^1$ is realized as a $3\times3$–BN–Sigmoid block. \\

\noindent Finally, a CBAM refresher consists of sequential channel and spatial attention modules, which focus on the channel and spatial dimensions, respectively. This mechanism recalibrates feature responses, while a residual connection preserves the original semantics, similar to the attention strategy employed in \cite{Bui2023}.:
\begin{equation}
\tilde{F}_i = \mathrm{CBAM}\bigl(Z'_i\bigr) + F_i.
\end{equation}

\noindent The output $\tilde{F}_i\in\mathbb{R}^{B\times C_i\times H_i\times W_i}$ proceeds to the next Up‐block and the local classifier $\pi_i$. Compared to Laplacian‐based EGA, the only learnable parameters here lie in $\phi$, $\psi$, and CBAM, while the wavelet extractor remains parameter‐free.

\subsection{Objective Function}
\noindent Following prior work in medical image segmentation, we train MEGANet-W with a deep‐supervised hybrid loss that balances pixel‐wise accuracy and region overlap. Concretely, each side‐output \(D_k\) is compared with its matched ground‐truth mask \(g_k\) via the sum of binary cross‐entropy and soft Dice:

\begin{flalign}\label{eq:bce-dice}
&\mathcal{L}_{\mathrm{BCE+Dice}}\bigl(P_k, g_k\bigr) &&\notag\\
&\quad= -\frac{1}{N}\sum_{i=1}^N 
     \Bigl[g_{k,i}\log P_{k,i} + (1-g_{k,i})\log\bigl(1-P_{k,i}\bigr)\Bigr] &&\notag\\
&\quad\quad + 1
 - \frac{2\sum_{i=1}^N P_{k,i}\,g_{k,i}}
         {\sum_{i=1}^N P_{k,i} \;+\;\sum_{i=1}^N g_{k,i} \;+\;1}\,.
\end{flalign}

\noindent The final objective is the unweighted sum over all scales,
\begin{equation}
\mathcal{L}
=\sum_{k=1}^{D}\mathcal{L}_{\mathrm{BCE+Dice}}\bigl(P_k, g_k\bigr)\,.
\end{equation}

\noindent Here \(D\) is the number of decoder layers (we set \(D=5\) as in U-Net). \(P_k\) denotes the predicted feature map at the \(k\)th decoding layer, and \(g_k\) is the ground-truth polyp mask at scale \(k\).

\begin{table*}[!t]
    \centering
    \scriptsize
    \caption{%
        \textbf{Quantitative results.} Performance of various polyp segmentation methods on five polyp segmentation datasets. Metrics shown are mean Intersection over Union (mIoU), mean Dice (mDice), and Mean Absolute Error (MAE).
    }
    \label{tab:quantitative}
    \resizebox{\textwidth}{!}{%
    \begin{tabular}{l|ccc|ccc|ccc|ccc|ccc}
    \toprule
    \textbf{Methods}
      & \multicolumn{3}{c|}{\textbf{CVC-300}}
      & \multicolumn{3}{c|}{\textbf{ClinicDB}}
      & \multicolumn{3}{c|}{\textbf{Kvasir-SEG}}
      & \multicolumn{3}{c|}{\textbf{ColonDB}}
      & \multicolumn{3}{c}{\textbf{ETIS}} \\
    \cmidrule(lr){2-4} \cmidrule(lr){5-7} \cmidrule(lr){8-10} \cmidrule(lr){11-13} \cmidrule(lr){14-16}
     & mIoU & mDice & MAE
     & mIoU & mDice & MAE
     & mIoU & mDice & MAE
     & mIoU & mDice & MAE
     & mIoU & mDice & MAE \\
    \midrule
    U-Net\cite{Ronneberger2015}          & 62.7 & 71.0  & 2.2  & 75.5  & 82.3 & 1.9 & 74.6  & 81.8  & 5.5  & 44.4  & 51.2 & 6.1  & 33.5  & 39.8 & 3.6  \\
    U-Net++\cite{Zhou2018}        & 62.4 & 70.7  & 1.8  & 72.9  & 79.4 & 2.2 & 74.3  & 82.1  & 4.8  & 41.0  & 48.3 & 6.4  & 34.4  & 40.1 & 3.5  \\
    SFA\cite{FangYuqiandChen2019}            & 32.9 & 46.7  & 6.5  & 60.7  & 70.0 & 4.2 & 61.1  & 72.3  & 7.5  & 34.7  & 46.9 & 9.4  & 21.7  & 29.7 & 10.9 \\
    PraNet\cite{Fan2020}         & 79.7 & 87.1  & 1.0  & 84.9  & 89.9 & \underline{0.9} & 84.0  & 89.8  & 3.0  & 64.0  & 70.9 & 4.5  & 56.7  & 62.8 & 3.1  \\
    MSNet\cite{Zhao2021}          & \underline{80.7} & 86.9  & 1.0
                   & \underline{87.9} & \underline{92.1} & \textbf{0.8}
                   & \underline{86.2} & 90.7 & 2.8
                   & 67.8 & 75.5 & 4.1
                   & 66.4 & 71.9 & 2.0 \\
    PEFNet\cite{Nguyen-Mau2023}         & 79.7 & 87.1  & 1.0  & 81.4  & 86.6 & 1.0 & 83.3  & 89.2  & 2.9  & 63.8  & 71.0 & \textbf{3.6}  & 57.2  & 63.6 & \underline{1.9}  \\
    MEGANet (R-34)\cite{Bui2023} & 80.5 & \underline{88.3}  & \underline{0.9}
                   & \textbf{88.5} & \textbf{93.0} & \textbf{0.8}
                   & 85.9 & \underline{91.1} & \underline{2.7}
                   & \underline{70.6} & \underline{78.1} & \underline{3.8}
                   & \underline{67.9} & \underline{76.3} & 2.2 \\
    \textbf{Ours}  & \textbf{82.8} & \textbf{89.5} & \textbf{0.7}
                   & 87.3  & 91.8  & \textbf{0.8}
                   & \textbf{87.0} & \textbf{92.0} & \textbf{2.5}
                   & \textbf{70.7} & \textbf{78.7} & \textbf{3.6}
                   & \textbf{69.6} & \textbf{77.5} & \textbf{1.5} \\
    \bottomrule
    \end{tabular}%
    }
    \vspace{1pt}
    \makebox[\textwidth][l]{\scriptsize * Reproduced using official code}
\end{table*}

\section{Experiments}

\subsection{Datasets}
\noindent We evaluate our proposed \textbf{MEGANet-W} on five standard benchmark datasets: Kvasir-SEG \cite{Jha2019}, CVC-ClinicDB \cite{Bernal2015}, CVC-ColonDB \cite{Tajbakhsh2016}, ETIS \cite{Silva2014}, and EndoScene \cite{Vzquez2017}. Notably, the EndoScene \cite{Vzquez2017} dataset comprises 912 images from two subsets: CVC-ClinicDB and CVC-300. To ensure a fair comparison, we adopt the same experimental setup as MEGANet, selecting 1,450 images from Kvasir-SEG (900 images) and CVC-ClinicDB.

\subsection{Implementation Details}
\noindent MEGANet-W is implemented in PyTorch~2.0 and trained for 200 epochs on 1,450 images using a single NVIDIA Tesla P100 (16\,GB). We use SGD (momentum 0.867472, weight decay $3.5454\times10^{-6}$) with batch size 16. Inputs are resized to $416\times416$ (upsampled for evaluation), and data augmentation includes random flips, rotations, color jitter and multi-scale training ($\{0.75,1.00,1.25\}$). The checkpoint with lowest training loss is selected for final evaluation on five held‐out datasets (total runtime $\approx10.5$\,h).

\subsection{Evaluation Metrics}

\noindent To quantify the performance of the proposed MEGANet-W and enable direct comparison with other approaches, we utilize five standard metrics: the mean Dice coefficient (mDice), the mean Intersection over Union (mIoU), accuracy, precision, and recall. Each metric is computed from the counts of true positives (TP), true negatives (TN), false positives (FP), and false negatives (FN) as follows:

\begin{align}
\mathrm{mDice}    &= \frac{2\,\mathrm{TP}}{2\,\mathrm{TP} + \mathrm{FP} + \mathrm{FN}},\\
\mathrm{mIoU}     &= \frac{\mathrm{TP}}{\mathrm{TP} + \mathrm{FP} + \mathrm{FN}},\\
\mathrm{Accuracy} &= \frac{\mathrm{TP} + \mathrm{TN}}{\mathrm{TP} + \mathrm{TN} + \mathrm{FP} + \mathrm{FN}},\\
\mathrm{Precision}&= \frac{\mathrm{TP}}{\mathrm{TP} + \mathrm{FP}},\\
\mathrm{Recall}   &= \frac{\mathrm{TP}}{\mathrm{TP} + \mathrm{FN}}.
\end{align}

\noindent These metrics were computed across five widely‐used colonoscopy datasets: \textit{CVC‐300}, \textit{CVC‐ClinicDB}, \textit{Kvasir}, \textit{CVC‐ColonDB}, and \textit{ETIS‐LaribPolypDB}, allowing us to evaluate the robustness and generalizability of the proposed model across diverse image characteristics and dataset distributions.

\section{Results}
\noindent We quantitatively compare our \textbf{MEGANet-W} against seven state-of-the-art (SOTA) methods from the original MEGANet baseline paper: U-Net \cite{Ronneberger2015}, U-Net++ \cite{Zhou2018}, SFA \cite{FangYuqiandChen2019}, PraNet \cite{Fan2020}, MSNet \cite{Zhao2021}, and PEFNet \cite{Nguyen-Mau2023}, as detailed in Sec.\ref{sec:quantitative_result}. The results for MEGANet \cite{Bui2023} were obtained by rerunning the model locally, while the results for all other methods are reported directly from the MEGANet paper. We also qualitatively compare our \textbf{MEGANet-W} with the original MEGANet baseline in Sec.\ref{sec:qualitative_result}.

\subsection{Quantitative Results}
\label{sec:quantitative_result}
\noindent We evaluate MEGANet-W against a range of baselines. As shown in Table \ref{tab:quantitative}, our method consistently achieves the best or second-best performance across all datasets in terms of mIoU, mDice, and MAE. Notably, MEGANet-W outperforms all competing methods on CVC-300 with an mIoU of 82.8\% and mDice of 89.5\%, while also delivering SOTA results on Kvasir-SEG (87.0\% mIoU, 92.0\% mDice) and ETIS (69.6\% mIoU, 77.5\% mDice), confirming its robustness on both standard and challenging datasets. Across all datasets, MEGANet-W delivers the best MAE values, indicating improved pixel-level precision and better boundary localization. These results confirm the effectiveness of incorporating wavelet-based edge guidance within the decoder attention mechanism.

\subsection{Qualitative Results}
\label{sec:qualitative_result}

\begin{figure}[t]
  \centering
  \footnotesize               
  \setlength{\tabcolsep}{1pt}
  \resizebox{\columnwidth}{!}{%
    \begin{tabular}{cccc}
      \textbf{Input} & \textbf{GT} & \textbf{Baseline} & \textbf{Ours} \\[2pt]
      \includegraphics[width=0.17\columnwidth]{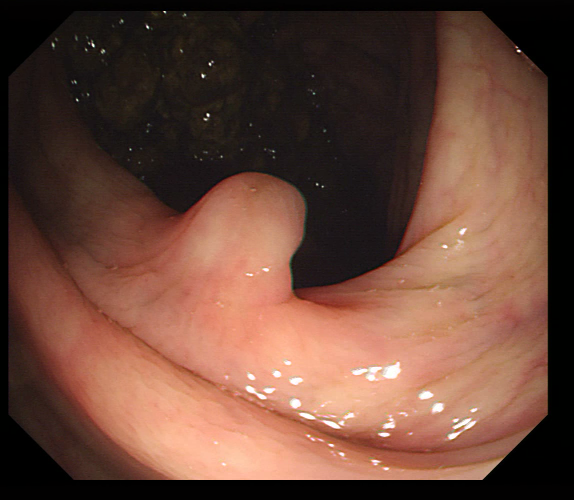}      &
      \includegraphics[width=0.17\columnwidth]{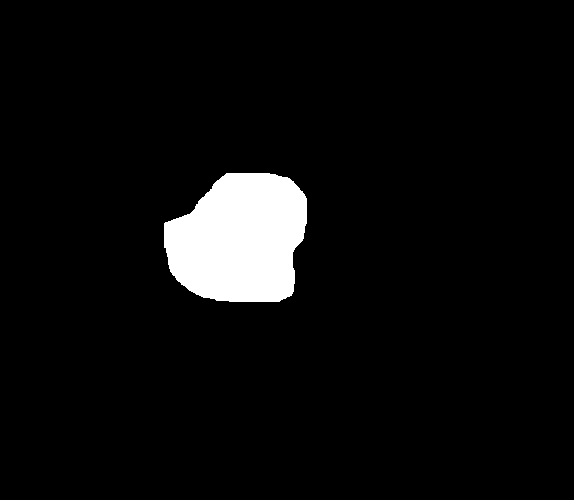}   &
      \includegraphics[width=0.17\columnwidth]{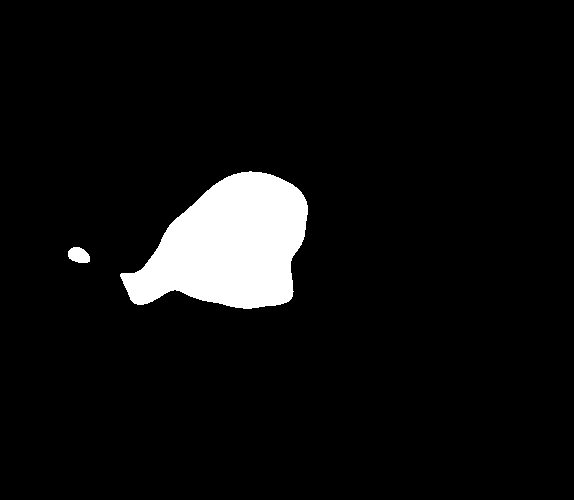}  &
      \includegraphics[width=0.17\columnwidth]{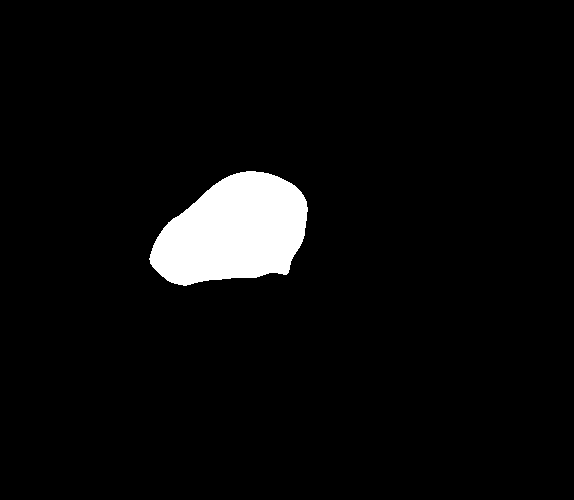} \\[2pt]
      \includegraphics[width=0.17\columnwidth]{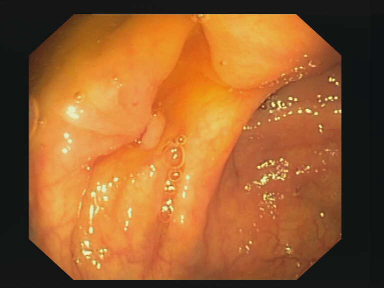}      &
      \includegraphics[width=0.17\columnwidth]{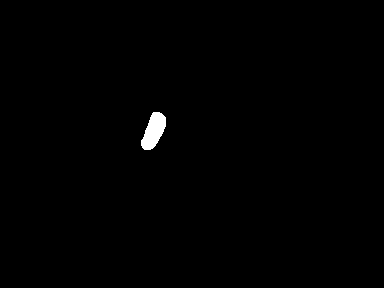}   &
      \includegraphics[width=0.17\columnwidth]{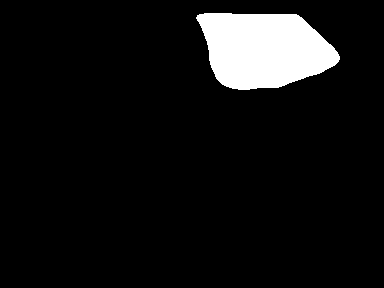}  &
      \includegraphics[width=0.17\columnwidth]{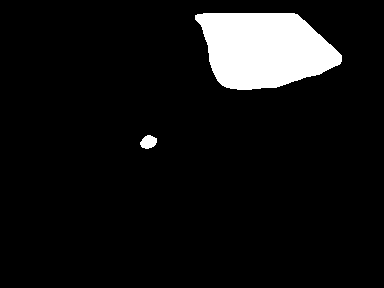} \\[2pt]
      \includegraphics[width=0.17\columnwidth]{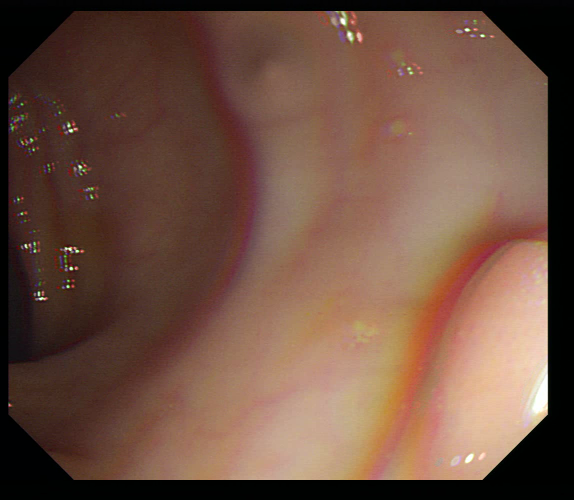}      &
      \includegraphics[width=0.17\columnwidth]{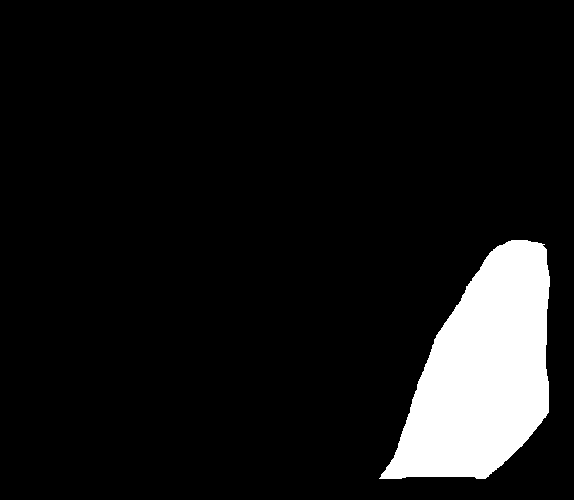}   &
      \includegraphics[width=0.17\columnwidth]{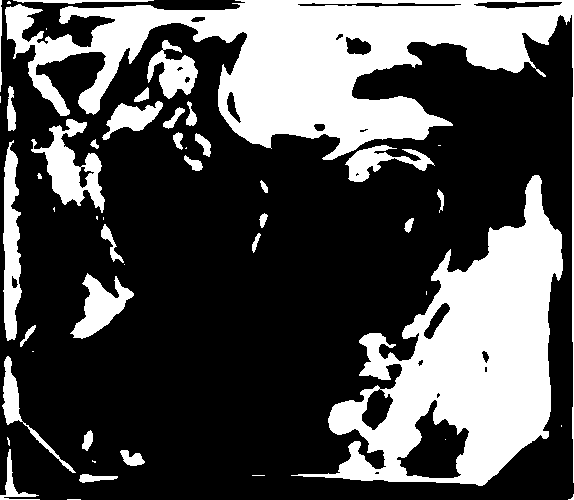}  &
      \includegraphics[width=0.17\columnwidth]{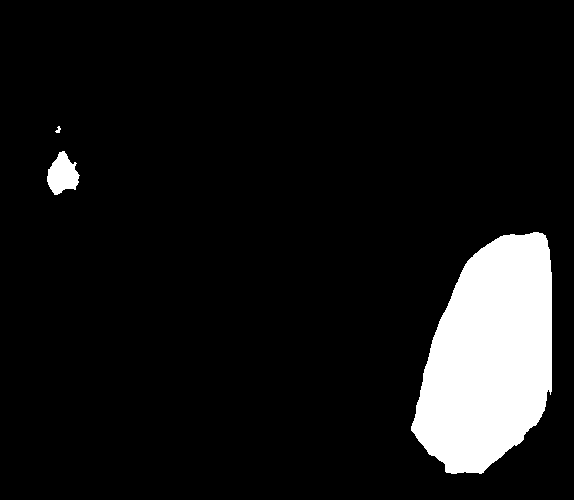} \\[2pt]
      \includegraphics[width=0.17\columnwidth]{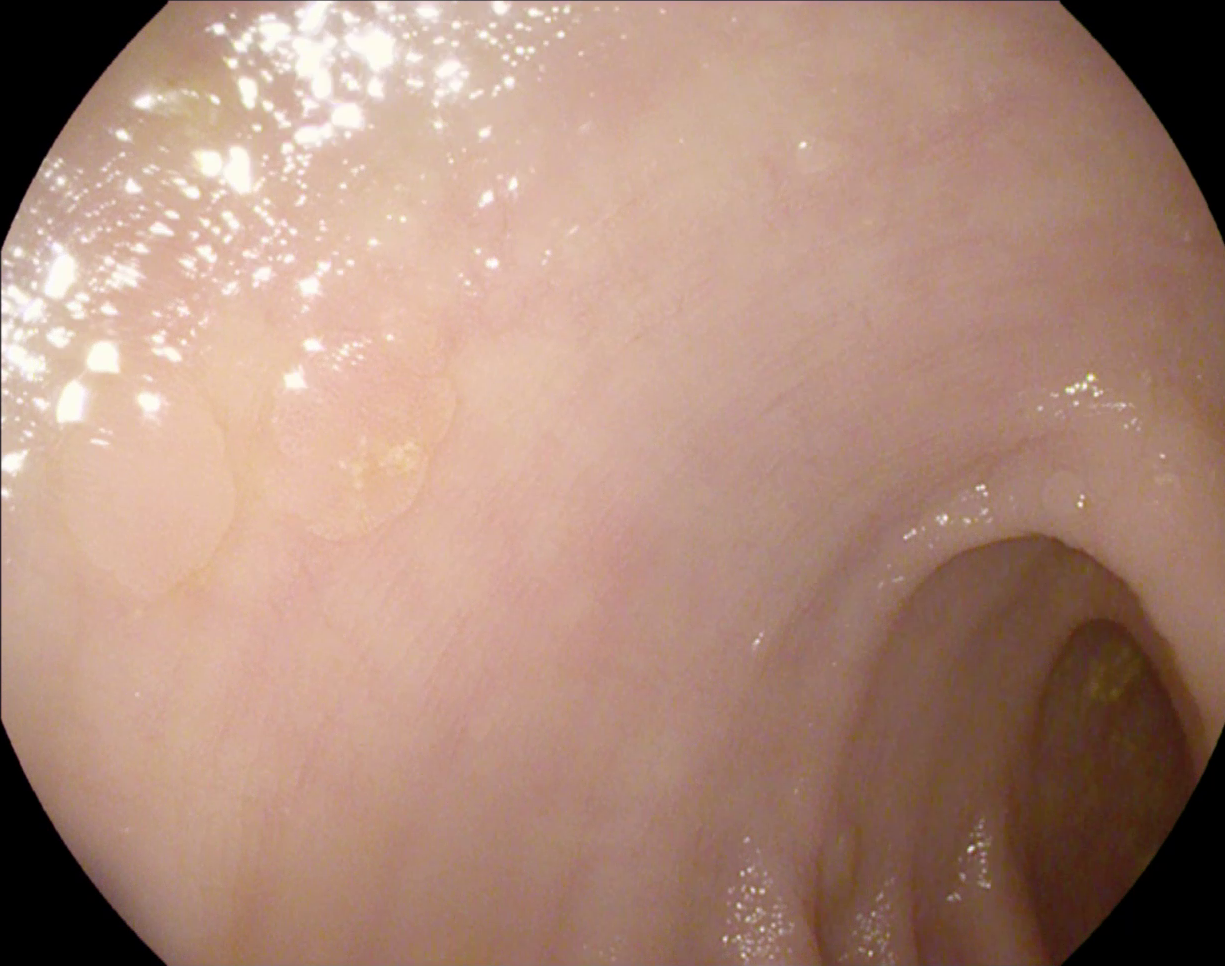}      &
      \includegraphics[width=0.17\columnwidth]{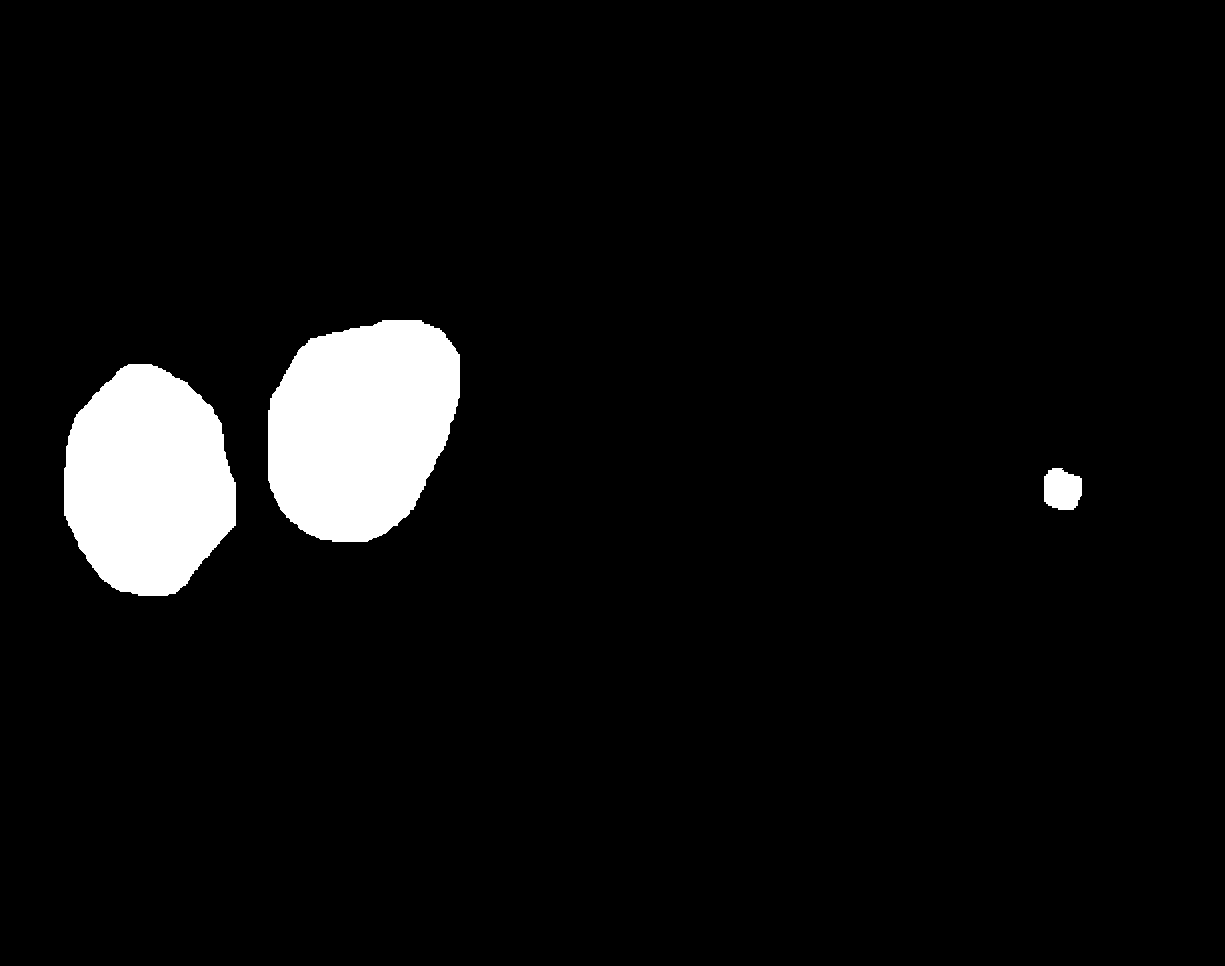}   &
      \includegraphics[width=0.17\columnwidth]{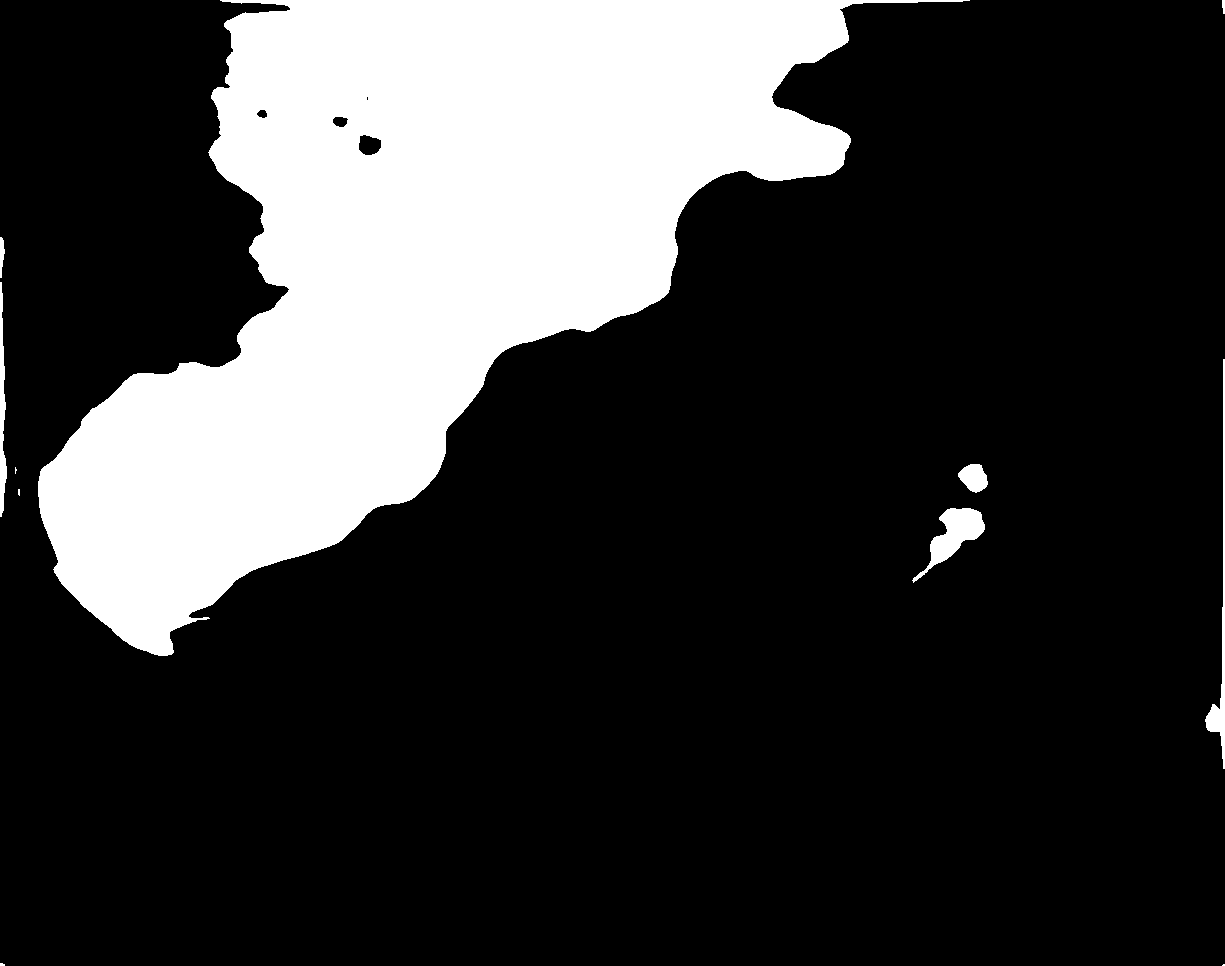}  &
      \includegraphics[width=0.17\columnwidth]{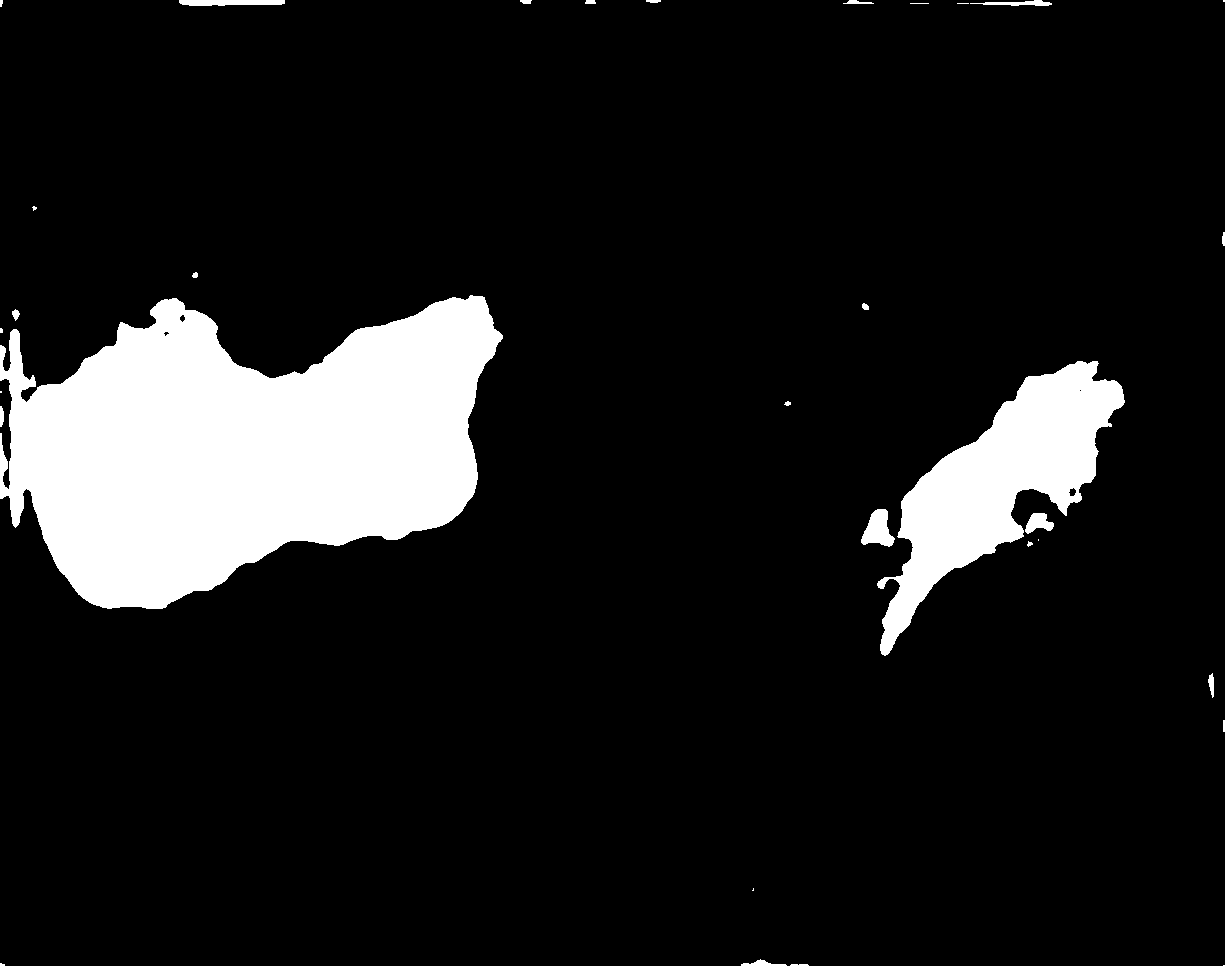} \\[2pt]
      \includegraphics[width=0.17\columnwidth]{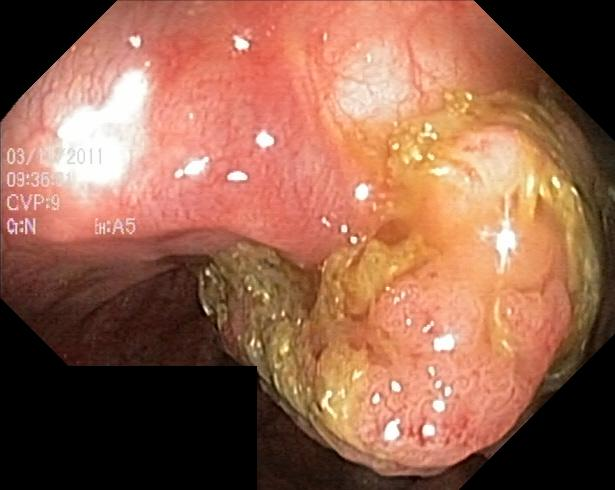}      &
      \includegraphics[width=0.17\columnwidth]{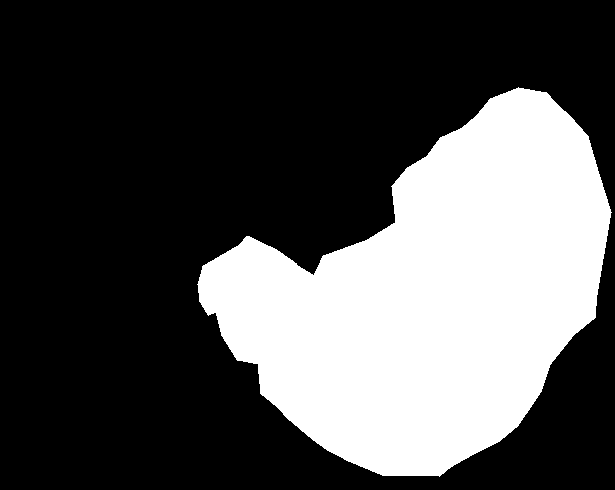}   &
      \includegraphics[width=0.17\columnwidth]{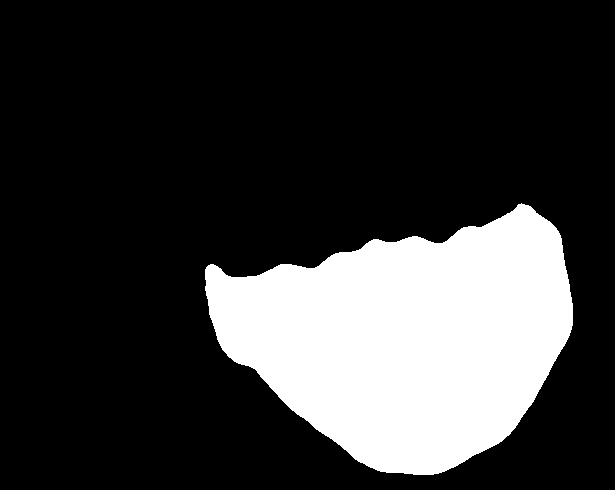}  &
      \includegraphics[width=0.17\columnwidth]{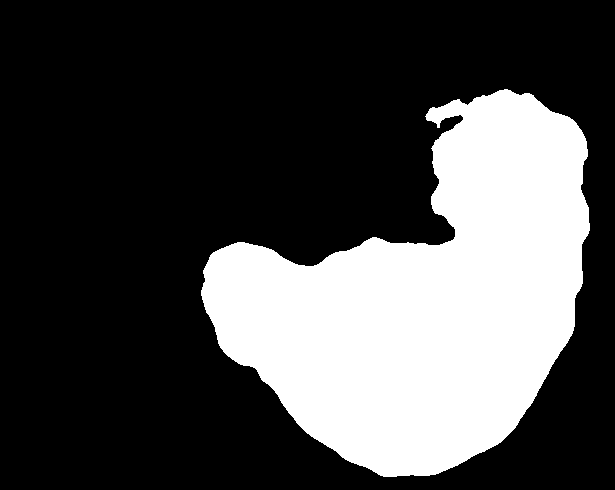} \\
    \end{tabular}%
  }
  \caption{\textbf{Qualitative results.} Each row shows one test sample drawn from a different dataset: CVC-300, CVC-ClinicDB, CVC-ColonDB, ETIS-LaribPolypDB, and Kvasir. Columns are (1) input frame, (2) ground-truth mask, (3) MEGANet-R34 (baseline), (4) MEGANet-W (ours).}
  \label{fig:qualitative}
\end{figure}

\noindent To further demonstrate the effectiveness of our proposed MEGANet-W, we present qualitative comparisons with the baseline MEGANet \cite{Bui2023} on representative samples that present different segmentation challenges, as shown in Fig.\ref{fig:qualitative}. Each row highlights a different difficulty (i.e., low contrast, background similarity, illumination artifacts, or morphological complexity), together with the corresponding ground-truth masks and predicted outputs. \\

\noindent In the first row, the polyp is difficult to localize because of the low contrast with the surrounding tissue. While both models detect the region, MEGANet-W provides more precise boundary delineation. The second row presents a polyp with color and texture similar to the background, where MEGANet-W offers better localization than the baseline, although the prediction is slightly affected by background noise. In the third row, strong specular highlights and illumination noise lead the baseline to generate scattered false positives, whereas MEGANet-W maintains a clean and accurate segmentation. The fourth row involves small, spatially separated polyps; MEGANet-R34 over-segments the region by merging background areas, while MEGANet-W correctly identifies and segments each polyp without merging errors. Finally, in the fifth row, the polyp exhibits an irregular shape and complex texture. MEGANet-W able to capture the structure more accurately and preserves boundary integrity better than the baseline.

\section{Conclusion and Future Works}
\noindent In this work, we presented MEGANet-W for weak-boundary polyp segmentation. Building upon the strong foundation of the original MEGANet architecture, we integrate a parameter-free, two-level Haar wavelet head into each decoder stage and W-EGA module to fuse multi-orientation wavelet edges with reverse-attention and input-edge branches. MEGANet-W achieves more precise boundary delineation without increasing model complexity and consistently outperforms SOTA methods across five public polyp datasets in terms of mIoU, mDice, and MAE. Looking forward, futher work can focus on extending our work to video colonoscopy or volumetric data, paving the way for real-time tracking and 3D segmentation. Additionally, validating MEGANet-W on diverse, multi-center clinical datasets and enhancing its interpretability through visualization of wavelet attention responses will be essential steps toward clinical deployment and trust.

\section*{Acknowledgements}
\noindent We appreciate Xin Ying, Zhe Enn, and Zhe Yang for providing the computational units used in this work. We also thank Leon Loo and Zheng Yeh for their support. This research work is supported by Xiamen University Malaysia Research Fund (Grant No: XMUMRF/2024-C14/IECE/0055).

\bibliographystyle{IEEEtran}
\bibliography{cvProject}

\end{document}